# Pressure Model of Soft Body Simulation

Maciej Matyka*  
University of Wrocław, Poland

Mark Ollila [†]  
Linköping University, Sweden

**Abstract**

Motivated by existing models used for soft body simulation which are rather complex to implement, we present a novel technique which is based on simple laws of physics and gives high quality results in real-time. We base the implementation on simple thermodynamics laws and use the Clausius-Clapeyron state equation for pressure calculation. In addition, this provides us with a pressure force that is accumulated into a force accumulator of a 3D mesh object by using an existing spring-mass engine. Finally after integration of Newtons second law we obtain the behavior of a soft body with fixed or non-fixed air pressure inside of it.

**CR Categories:** I.6.8 [Simulation and Modeling]: Types of Simulation—Animation; I.3.5 [Computer Graphics]: Computational Geometry and Object Modeling—Physically based modeling I.3.7 [Computer Graphics]: Three-Dimensional Graphics and Realism—Animation

## 1 Introduction and Background

In this paper we present a model for three dimensional deformable objects simulation based on simple and fundamental physics principles.

We started our research in deformable object animation because of the complexity and cost of solutions with FEM, FEV (see [1; 15]) and LEM (see [9; 10; 11; 12]) methods. Some attempts with finite elements in real time simulations have been made (see [2]), however, the complexity of given algorithms is still high. Also similar problems appear with the Green function approach (see [6]) - complexity of implementation is high. Fast realtime pre-computed data-driven models of interactive physically based deformable scenes were proposed also in [14]. Methods using Navier-Stokes equations for Soft Bodies have been presented in [3]. The authors use Navier-Stokes equations for compressible fluid to compute properties of compressible fluid enclosed in a mesh. The model gives good soft behavior but solution speed is not efficient to achieve good results for real time animation.

In this paper, we introduce a novel idea of using ideal gas law in calculating pressure force. For pressure calculation we use a simple ideal gas state equation. Using the ideal gas approximation results in fast and physically correct animations.

*e-mail:maq@panoramix.ift.uni.wroc.pl  
[†]e-mail:marol@itn.liu.se

The model which we introduce, is in a fact described by two equations and be very easy implemented into existing spring-mass systems by adding one additional force - the pressure force.

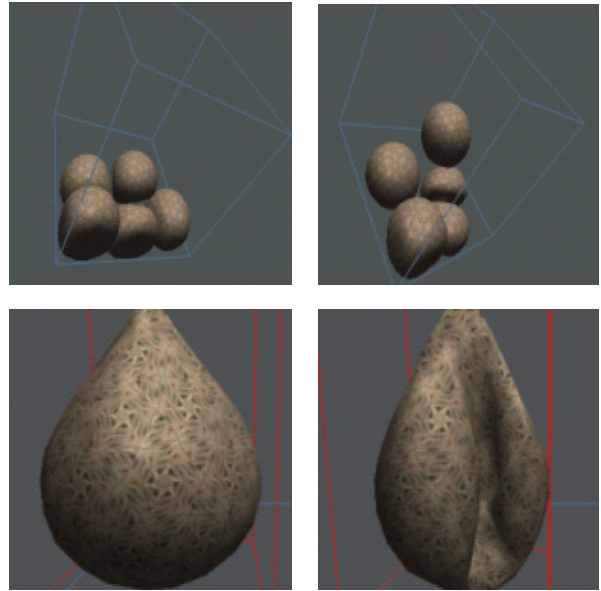

Figure 1: Real time animations created using presented model. Each ball object contains: 320 faces (162 vertices). Both examples run realtime on 800Mhz Duron processor machine (approximately 40-50fps, depending on collision detection configuration).

### 1.1 Particle System

Let us consider the governing elements of simple Spring-Mass (SM) model (see [4]. We will show how to expand it to simulate a 3D soft body with deformations. A simple SM engine contains a couple of obvious techniques, with the most important features to work with is a simple particle system which uses simple physics (Newton Laws), compute forces (Gravity, Hooke's linear spring force with damping) and makes numerical integration of one equation[1]:

$$\vec{F}_i = m_i \cdot \frac{\partial^2 \vec{r}_i}{\partial t^2} \quad (1)$$

## 2 Pressure Based Method for Soft Bodies

### 2.1 Method

The method presented in this paper is a soft body model based on classic cloth simulation (see [13]). Applying a wind force to cloth simulation results in very nice, good looking behavior.

---
[1] Where *i* indexes particles.

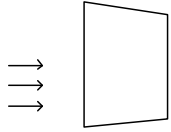

Figure 2: Cloth rectangle with fixed edges.

Lets say that we are presented with the situation depicted in figure (2). A cloth rectangle with fixed edges is placed, where wind, a force vector, is normal to initial surface of rectangle. What we obtain from the simulation like this is deformation of cloth under wind force.

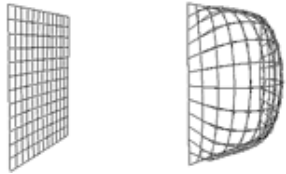

Figure 3: Cloth deformation under wind force.

Observing figure (3) it becomes apparent that it seems to be very similar to part of a deformed three dimensional object. Evolving the model further, we can "close the object" and put the "wind source" within it. Let us define model of shape as sketched in the figure (4).

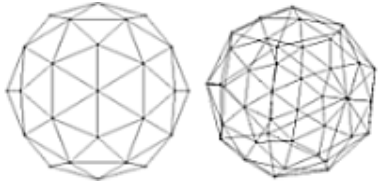

Figure 4: Triangular shape model with both solid and wireframe views presented.

Now let us imagine that we have some kind of "wind source" inside of that shape, which will introduce that nice deformation effect from figure (3) but in all directions. This is the basic idea of the implementation in this paper.

### 2.2 The Pressure Force

The simplest definition of pressure says that pressure is a force, acting on incremental surface elements and is parallel to normal vector of the surface. We would like to apply the force, which will act on our shape, and keep the shape geometry, but allow for possible deformations. Pressure, is a force that is always acting in a direction of normal vectors to the surface, so the shape will not deform significantly. So far we have only a definition of the pressure force vector, and we have to find some expressions for the force value. The expression for pressure in a specified point in the space acting on a surface is given by:

$$\vec{P} = P \cdot \hat{n} \left[ \frac{N}{m^2} \right], \qquad (2)$$

where $P$ is a pressure value and $\hat{n}$ is normal vector to surface on which pressure force is acting. For calculating pressure force we have to multiply $\vec{P}$ by $A[m^2]$ - the area of the surface. That gives us pressure force expression [2]:

$$\vec{F_P} = \vec{P} \cdot A\,[N], \qquad (3)$$

Now we will explain how to calculate $P$ - the pressure force value.

### 2.3 Ideal Gas Approximation

In our model, we will use thermodynamic approximation known as "Ideal Gas Approximation" (see [5] for detailed description of that approximation). We can use this approximation because our interest is more in the macroscopic level effects of gas presence. At this level we can assume that in an object a gas of particles without interactions exist. We are only interested in the statistical properties (i.e. average momentum given from particles to the model surfaces in a specified incremental time).

The ideal gas approximation gives us simple relationship between pressure value, temperature of gas, and macroscopic volume of the body which can be expressed i.e. by the well known Clausius Clapeyron equation:

$$PV = nRT, \qquad (4)$$

where $P$ is pressure value, $V$ is volume of the body, $n$ is gas mol number, $R$ is the ideal gas constant ($R = N_a k_b$, $N$ - Avogardo number, $k_b$ - Boltzmann constants), $T$ is a temperature. From equation (4) we can easy get an expression for pressure if we know values of temperature and volume of the body:

$$P = V^{-1} nRT, \qquad (5)$$

In the model presented in this paper we assume that $T = const$ and only the volume of a soft body changes. Specified assumptions will give us a very clear and easy to implement algorithm of pressure force calculation.

### 2.4 Algorithm

Before we dive into implementation details we will show the general algorithm of the presented solution. It will help clarify specific problems which appear during the implementation. The algorithm is based upon an existing particle spring-mass system with one modification - addition of a pressure force calculation. One computational step of the algorithm is as follows:

1 Loop over all particles:

    [1.1] Calculate and accumulate gravity and spring forces for all particles.

2 Calculate volume of the soft body

3 Loop over all faces:

    [3.1] Calculate pressure force acting on the face

    [3.1.1] Calculate field of the face

    [3.1.2] Calculate the pressure value

    [3.1.3] Loop over particles which define the face:

    [3.1.3.] Multiply result by field of the face and $\hat{n}$

    [3.1.3.2] Accumulate finally pressure force to the particle.

4 Integrate momentum equation

5 Resolve collision handling and response.

6 Move particles

---
[2] Where [N] means simply a force dimension.

# 3 Implementation

## 3.1 Calculate volume of the soft body

In Step 2 of the presented algorithm we have to calculate volume of the body. In the presented solution we used a variety of bounding objects to approximate that value. The implementation is well suited for non-complicated objects, and we will discuss later what kind of improvement can be done there. Bounding objects are a well known technique to speed up collision detection systems, and here we have used it to compute volumes. The type of bounding object depends strongly on geometry of simulated soft body (i.e. it is not very good to approximate ball by simple bounding box). In the model presented here we implemented three different bounding objects: bounding box, bounding ball, and bounding ellipsoid.

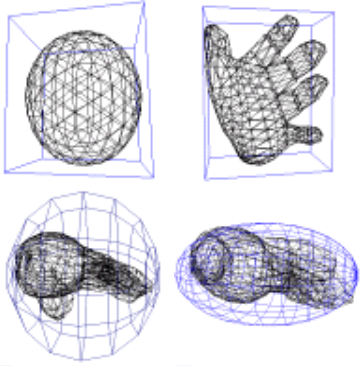

Figure 5: Three different types of bounding objects applied to the model.

We know calculating of volumes for bounding boxes, spheres and ellipsoids is fairly easy. For example, for the ellipsoid with $r_x$, $r_y$ and $r_z$ radiuses we have the expression for volume:

$$V_{el} = \left(\frac{4}{3}\right) \cdot \pi \cdot r_x r_y r_z \quad (6)$$

We use the term "bounding volume" for this technique of volume generation. Bounding volumes are not very accurate, but for a model, only the general change of body volume is needed. Of course for the hand model presented in the figure (5) a better approximation has been made with ellipsoid, and generally the ellipsoid has the best properties and is the most usable as a bounding volume.

## 3.2 Face field computation

Face field computation is quite simple, especially because of triangulated objects which we use as a models in the simulator. Simple algebraic vector multiplication of two edges is used here.

## 3.3 Numerical Methods

In the presented solution no special focus on numerical method has been done. We have used explicit Euler, Midpoint and RKIV integrators to integrate motion equations for every particle. It appears, that best choice is to use Mid Point algorithm, since we found that for some parameters configuration model is stable. All results presented have been calculated using $2^{nd}$ Mid Point scheme.

## 3.4 Summary of the implementation

Calculation of the body volume is one of the most important. After obtaining body volume, and the face field, we are able to calculate pressure force vector. Then basically accumulation of this force is performed.[3]

## 3.5 Collision Detection and Response

Existing techniques for collision detection and response (see [7; 8]) for deformable objects could be applied to described model of soft body simulation. For our purposes, for collision detection we use simple techniques of collisions with bounding objects. It is big simplification, but works very well for objects such as balls, cones, boxes, and other which are similar to the bounding objects. However, It is not very accurate, and future research needs to be performed here. Fortunately the results are rather good with the sample models used here.

The algorithm first detects which objects in the scene can collide. In the presented solution we used simple bounding boxes for first step of detection of possible contact.

Then a simple test of each point - bounding (ellipsoid/ball/box/other) object is applied. For each implicit formulation of bounding object, a simple value of the function is to be computed with the tested point. Tests have to be done for every point of one soft body with bounding objects of another soft body[4]. After that, if a point collides, we cut off $d\vec{r}$, the vector of movement (which obviously is to be given by the integration function to collision handling procedure). We have also implemented a simple iterative procedure which corrects the particle position if a particle intersects. That procedure checks if the point is inside of bounding object of the other soft body, if yes, then iteratively moves that point outside of the body as long as it will be outside of that bounding object.

When we recognize a collision, we implement a fairly simple collision response procedure. We divide the velocity vector of the particle into two velocities (parallel and tangent to collision surface) and we reflect the tangent part of it. Then both parts of velocity vector are multiplied by scalar coefficient to get effects of energy lost of particle during collision.

# 4 Results

In this section we present the results of the working application which was based on presented algorithm. In figures (1), (8) and (7) examples of working application are shown. All of these are taken directly from real time working animations. In the first two figures five bouncing balls with collisions, are placed in rotating box. The hanging ball with different pressures is shown in the next two figures. One ball object in the simulations contains 162 vertices and 320 faces. The third example in the figure (8) is a result of simulation with user interaction, where user can hit the simulated soft body (a hand in this case) with the rigid ball. The hand object contains ≈370 vertices with 736 triangles implemented into existing particle spring-mass engine. The simulations were computed in real time and run at (40-50 fps) on a Duron 800Mhz with RivaTNT2 graphics card.

In the figure (6) we present a plot of time needed for computation of pressure force. We prepared tests for an Athlon XP 1.8Ghz processor. The sketch shows times in *ms* for several different number of particles in the simulation space and average - trend line. We did not apply collision detection during these tests.

---

[3]Please note that we have to apply force which is parallel to normal vector of the point (simple calculated by average of normal of its neighbor faces) and divide by integer value equal to number of faces, which contain that particle.

[4]Some tests can be done there, if checking first point-bounding box collision, then i.e. point-bounding ellipsoid is faster than checking only point-bounding ellipsoid intersection.

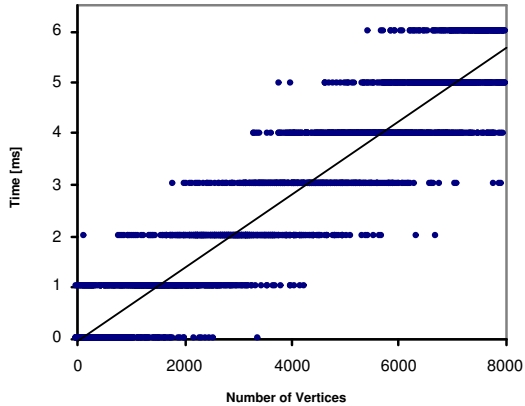

Figure 6: Time of pressure force calculation in *ms* in a function of number of vertices. Figure shows increasing of calculation cost, while number of vertices / triangles increases.

## 5  Conclusions and Future Work

Here we have presented a model on soft bodies using pressure. The most important advantages of the model is that it is fast (see figure 6), stable and can be easily implemented. However, there is significant future work to be performed. This includes work with more advanced volume calculation where we believe two methods are the most interesting to pursue. This includes development of some kind of bounding primitive subdivision algorithm to refine the volume representation. An alternative, is to compute volumes of the bodies with a Monte Carlo integration procedure. However, even though Monte Caro methods are accurate and converge to good results, they are rather slow and may not be suitable for real time applications. Future work will also include investigation into more complex objects and the suitability of this model with other forms of model representations such as Hypertextures.

## 6  Acknowledgements

The hand object presented in results section has been created by Mariusz Jarosz. Authors wish to also thank Jakub Kominiarczuk and Marcin Wojciechowski for support of the work. We are also grateful to an anonymous reviewer who motivated us to make the paper much better. The paper is the result of project work in the Modeling and Animation course at Linköping University.

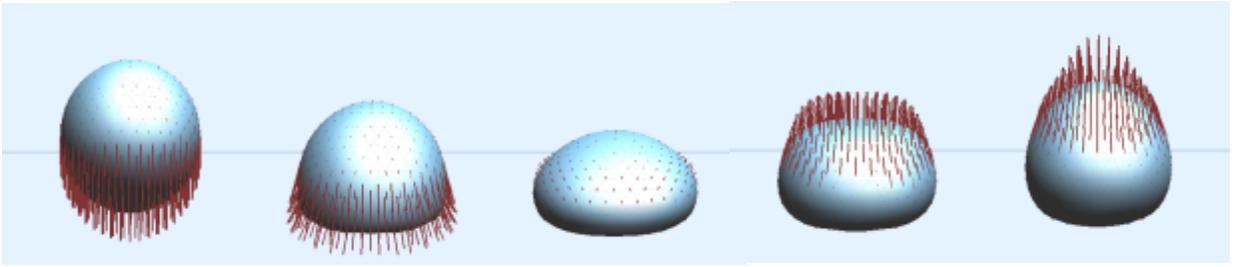

Figure 7: Bouncing ball with deformations. Visualization with velocity vectors for better information about physical property of moving object (i.e. momentum). Frame rate: 50fps. Details: 320 faces (162 vertices).

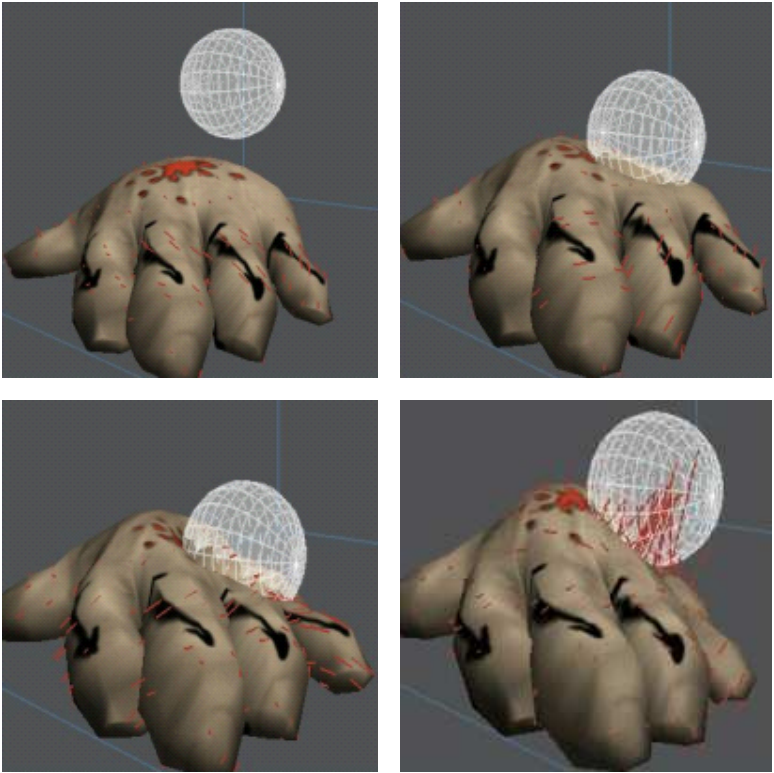

Figure 8: Hand object in gravity field with example of user interaction. Frame rate: 50fps. Details: 768 faces (386 vertices).